\documentclass{moriond}

\def\Journal#1#2#3#4{{#1} {\bf #2}, #3 (#4)}

\def\NIMA{{\em Nucl. Instrum. Methods} A}

\def\PRL{\em Phys. Rev. Lett.}
\def\PRD{{\em Phys. Rev.} D}

\def\CPC{{\em Chin. Phys.} C}
\def\EPJC{{\em Eur. Phys. J.} C}
\def\JHEP{{\em JHEP}}

\def\be{\begin{equation}}
\def\ee{\end{equation}}
\def\bea{\begin{eqnarray}}
\def\eea{\end{eqnarray}}

\begin{document}
\vspace*{4cm}
\title{Leptonic and semileptonic charm decays at BESIII}

\author{ChaoChen~\footnote{chaochen@ihep.ac.cn}\\(On behalf of the BESIII collaboration)}

\address{Institute of High Energy Physics, Chinese Academy of Sciences, Beijing, China}

\maketitle

\abstracts{
	The BESIII collaboration has achieved important measurements in charmed purely leptonic and semi-leptonic decays using data samples collected at center-of-mass energies of 3.773 GeV, 4.128-4.226 GeV, and 4.237-4.669 GeV. This proceeding presents recent BESIII results on charmed purely leptonic and semileptonic decays, including measurements of branching fractions, the Cabibbo-Kobayashi-Maskawa matrix elements $|V_{cs}|$ and $|V_{cd}|$, decay constants and form factors, as well as a test of the Lepton flavor universality.}

\section{Introduction}

The BESIII experiment~\cite{BESIII:2009fln} started data taking since 2008 from symmetric $e^+e^-$ collisions at the Beijing Electron Positron Collider II (BEPCII)~\cite{Yu:2016cof}.
Recently, BESIII has collected the largest $D\bar{D}$ dataset at the center-of-mass energy of 3.773 GeV with an integrated luminosity of 20.3 fb$^{-1}$~\cite{BESIII:2020nme}.
Additionally, the BESIII detector also with data taking at the center-of-mass energies of 4.128-4.226 GeV with the process $e^+e^- \to D_s^\pm D_s^{*\mp}$ and process $e^+e^- \to D_s^{*+}D_s^{*-}$ at $\sqrt{s}=4.237$–4.669 GeV to study the $D_s$ decays\footnote{Charge conjugation is implied throughout these proceedings, except where explicitly stated.}.
The measurements of charmed leptonic and semileptonic decay at BESIII are conducted with the double-tag method~\cite{MARK-III:1985hbd}.
Purely leptonic and semileptonic decays of charm mesons ($D$ mesons) provide a powerful probe for testing the Standard Model (SM) and searching for possible contributions from physics beyond the SM.
As a result, these channels are of particular interest for determining decay constants, form factors, and the Cabibbo-Kobayashi-Maskawa (CKM) matrix elements, as well as for examining the unitarity of the CKM matrix and lepton flavor universality (LFU).

\section{Leptonic Decay $D\to\ell^+\nu_{\ell}$}

In the SM, the partial width of the purely leptonic decays of $D$ mesons ($D^+_{(s)} \to \ell^+\nu_{\ell}$, $(\ell=\mu,\tau)$) can be expressed as~\cite{Silverman:1988gc}:
\begin{equation}
	\small
	\Gamma_{D_{(s)}^{+} \rightarrow \ell^{+} \nu_{\ell}} = \frac{G_F^{2}}{8 \pi} f_{D_{(s)}}^{2} |V_{cs(d)}|^{2} m_{D_{(s)}} m_{\ell}^{2} \left(1 - \frac{m_{\ell}^{2}}{m_{D_{(s)}}^{2}}\right)^{2},
	\label{eqution:leptoWidth}
\end{equation}
where $G_F$ is the Fermi coupling constant, $m_{D_{(s)}}$ is the mass of the $D_{(s)}$ meson, and $m_{\ell}$ is the mass of the charged lepton. 
Hence, we can determine the branching fraction and the product of the CKM matrix elements $|V_{cs(d)}|$ and the decay constants $f_{D_{(s)}}$ using the lifetime of $D_{(s)}$ ($\tau_{D_{(s)}}$) in the decay of $D^+_{(s)} \to \ell^+\nu_{\ell}, (\ell=e,\mu,\tau)$. These measurements can be used to test the SM and theoretical calculations from lattice quantum chromodynamics (LQCD).

\subsection{$D^+ \to \mu^+ \nu_{\mu}$}

\begin{figure}[htbp]
	\centering
	\includegraphics[width=0.4\linewidth]{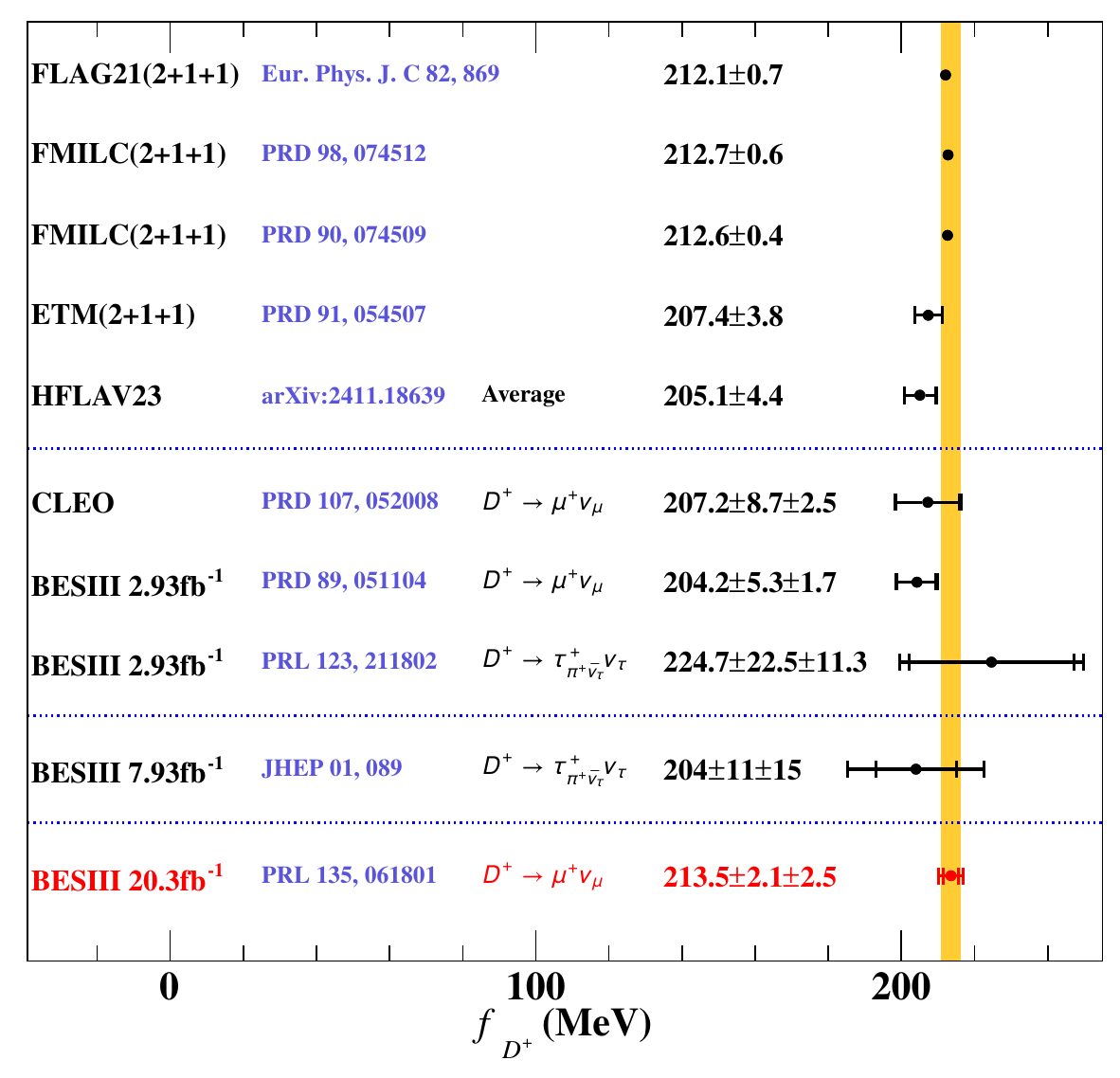}
	\includegraphics[width=0.4\linewidth]{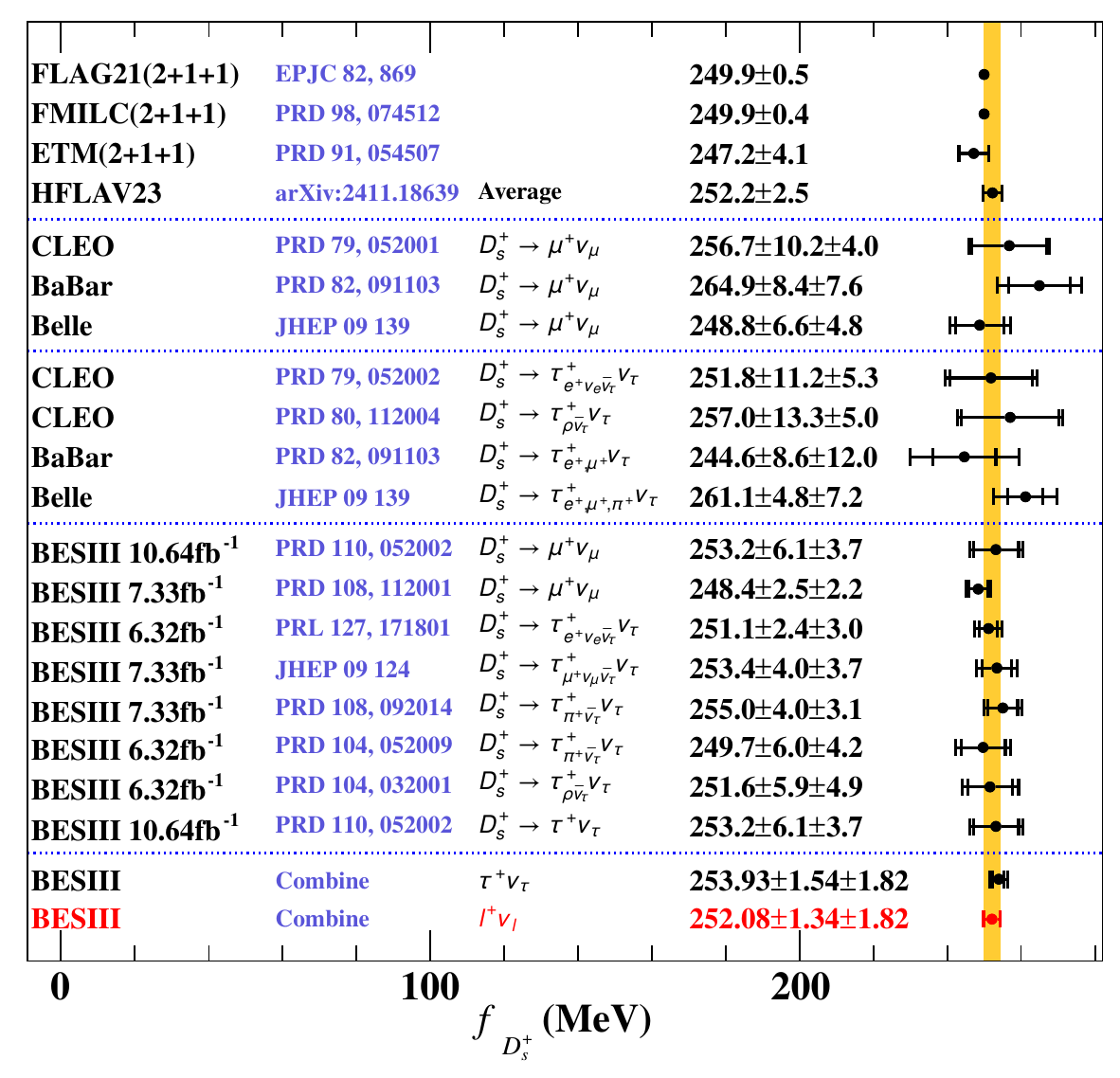}
	\caption{Comparison of the decay constants $f_{D^+}$ (left) and $f_{D_s^+}$ (right) from different theoretical predictions and experimental measurements.}
	\label{fig:decayconstant}
\end{figure}

The measurement of the purely leptonic decay $D^+ \to \mu^+ \nu_{\mu}$ is performed with a 20.3 fb$^{-1}$ dataset collected at a center-of-mass energy of 3.773 GeV~\cite{BESIII:2024kvt}. In contrast to previous analyses, his study incorporates the radiative correction factor $(1 + \frac{\alpha}{\pi} C_p)$, which includes structure-dependent bremsstrahlung, as well as both short- and long-distance electroweak corrections. The measured branching fraction for this decay is $(4.034 \pm 0.080_{\rm stat} \pm 0.040_{\rm syst}) \times 10^{-4}$, and $f_{D^+} |V_{cd}| = (48.02 \pm 0.48_{\rm stat} \pm 0.24_{\rm syst} \pm 0.12_{\rm ext})$ MeV with the $D^+$ meson lifetime $\tau_{D^+} = (1.033 \pm 0.005) \times 10^{-12}$ s~\cite{PDG:2024cfk}. The third uncertainty primarily originates from $\tau_{D^+}$. By combining the averaged decay constant $f_{D^+}$ from LQCD calculations~\cite{FLAG:2021npn}, the CKM matrix element $|V_{cd}| = 0.2265 \pm 0.0023_{\rm stat} \pm 0.0011_{\rm syst} \pm 0.0009_{\rm ext}$ is determined. Using the value of $|V_{cd}|$ from the global SM fit as input~\cite{PDG:2024cfk}, the decay constant $f_{D^+} = (213.5 \pm 2.1_{\rm stat} \pm 1.1_{\rm syst} \pm 0.8_{\rm ext})$ MeV is extracted, and this result is compared with previous experimental and theoretical values in Fig.~\ref{fig:decayconstant} (left).

\subsection{$D_s^+ \to \ell^+ \nu_{\ell}$ ($\ell=\mu, \tau$)}

Compared to previous measurements of the decays $D_s^+ \to \ell^+ \nu_{\ell}$ ($\ell = \mu, \tau$)~\cite{BESIII:2021anh,BESIII:2023cym,BESIII:2023fhe,BESIII:2021wwd,BESIII:2021bdp,BESIII:2023ukh} via $e^+ e^- \to D_s^{*-} D_s^+$, which used 7.33 fb$^{-1}$ of data at center-of-mass energies between 4.128 and 4.226 GeV, a recent simultaneous fit was performed using 10.64 fb$^{-1}$ of data collected at 4.237–4.699 GeV to constrain the four $\tau^+$ decay ($\tau^+ \to \pi^+ \bar{\nu}_{\tau}$, $\tau^+ \to \pi^+\pi^0 \bar{\nu}_{\tau}$, $\tau^+ \to e^+ \nu_e \bar{\nu}_{\tau}$, and $\tau^+ \to \mu^+ \nu_\mu \bar{\nu}_{\tau}$) channels for $D_s^+ \to \tau^+ \nu_{\tau}$~\cite{BESIII:2024dvk}. The branching fraction for $D_s^+ \to \tau^+ \nu_{\tau}$ is determined to be $(5.60 \pm 0.16_{\rm stat} \pm 0.20_{\rm syst})\%$, while for $D_s^+ \to \mu^+ \nu_{\mu}$, it is $(0.547 \pm 0.026_{\rm stat} \pm 0.016_{\rm syst})\%$ without assuming lepton flavor universality. The values of $f_{D_s^+}$ and $|V_{cs}|$ are obtained as $(252.1 \pm 1.3_{\rm stat} \pm 1.7_{\rm syst})$ MeV and $0.982 \pm 0.005_{\rm stat} \pm 0.007_{\rm syst}$, respectively, with correlated uncertainties taken into account. The result for $f_{D_s^+}$ is compared with theoretical predictions and experimental measurements in Fig.~\ref{fig:decayconstant} (right).

\section{Semi-leptonic decay}

In the SM, the semileptonic decays of $D_{(s)}$ mesons decay to the pseudoscalar ($P$), vector ($V$), axial-vector ($A$) and light scalar ($S$) meson are mediated by the flavor-changing $c \to q \ell^+ \nu_{\ell}$ transition ($q = s, d$). The corresponding transition matrix element, $\mathcal{M}$, between the initial and final meson states can be written as:
\begin{equation}
	\mathcal{M}(D_{(s)} \to P(V/A) \ell^+ \nu_{\ell}) = \frac{G_F}{\sqrt{2}} V_{cq} H^{\mu} L_{\mu},
\end{equation}
The leptonic current is given by $L_{\mu} = \bar{\nu}{\ell} \gamma{\mu} (1 - \gamma_5) \ell$, while the hadronic current is expressed as $H^{\mu} = \langle P(V/A) | \bar{q} \gamma_{\mu} (1 - \gamma_5) c | D_{(s)}(p_{D_{(s)}}) \rangle$.
Since the leptonic and hadronic currents do not interact via strong forces, the product of the hadronic form factors and the CKM matrix elements, $|V_{cq}|$, can be determined experimentally.

\subsection{$D \to P \ell^+ \nu_{\ell}$, ($P= \bar{K}$ or $\eta$)}

The $D \to \bar{K} \ell^+ \nu_{\ell}$ decay is an ideal process for determining the form factor $f_+^{D \to \bar{K}}(0)$ and the CKM matrix element $|V_{cs}|$. Using 20.3 fb$^{-1}$ of data at $\sqrt{s} = 3.773$ GeV, the branching fractions for $D^0 \to K^- e^+ \nu_e$, $D^0 \to K^- \mu^+ \nu_\mu$, $D^+ \to K^0 e^+ \nu_e$, and $D^+ \to K^0 \mu^+ \nu_\mu$ are measured with high precision. The branching fraction ratios $\mathcal{R}_{\mu/e}^{0} = 0.972 \pm 0.003{\rm stat} \pm 0.004_{\rm syst}$ and $\mathcal{R}_{\mu/e}^{+} = 0.982 \pm 0.004{\rm stat} \pm 0.002_{\rm syst}$ are in excellent agreement with the SM prediction of $0.975 \pm 0.001$, confirming LFU.
The product $f_+^{D \to \bar{K}}(0) |V_{cs}|$ is measured to be $0.7160 \pm 0.0007_{\rm stat} \pm 0.0014_{\rm syst}$. Using the PDG value of $|V_{cs}|$~\cite{PDG:2024cfk}, the form factor is determined to be $f_+^{D \to \bar{K}}(0) = 0.7355 \pm 0.0007_{\rm stat} \pm 0.0014_{\rm syst}$. Conversely, using lattice QCD calculations~\cite{FermilabLattice:2022gku}, $|V_{cs}|$ is determined to be $0.9608 \pm 0.0009_{\rm stat} \pm 0.0019_{\rm syst}$~\cite{BESIII:2026ydr,BESIII:2026uin}.

The forward-backward asymmetries of these four decays are determined for the first time in both the overall and different $q^2$ intervals, with results in agreement with theoretical predictions. A simultaneous fit to the measured partial decay rates and asymmetries of $D \to \bar{K} \ell^+ \nu_{\ell}$ searches for a scalar current contribution in the $c \to s \ell^+ \nu_\ell$ transition. No significant deviation from the SM is found in the positron channels, while a small non-zero $c_S^{\mu}$ is observed in the muon channels, with results ${\rm Re}(c_S^{\mu}) = 0.007 \pm 0.008_{\rm stat} \pm 0.006_{\rm syst}$ and ${\rm Im}(c_S^{\mu}) = \pm(0.070 \pm 0.013_{\rm stat} \pm 0.010_{\rm syst})$, corresponding to a $1.9\sigma$ deviation. The Wilson coefficients $c_R^{\ell}$ and $c_L^{\ell}$, constrained by combining $D_s^+ \to \ell^+ \nu_\ell$ decay rates, are consistent with zero~\cite{BESIII:2026ydr,BESIII:2026uin}.

\begin{figure}[htbp]
	\centering
	\includegraphics[width=0.9\linewidth]{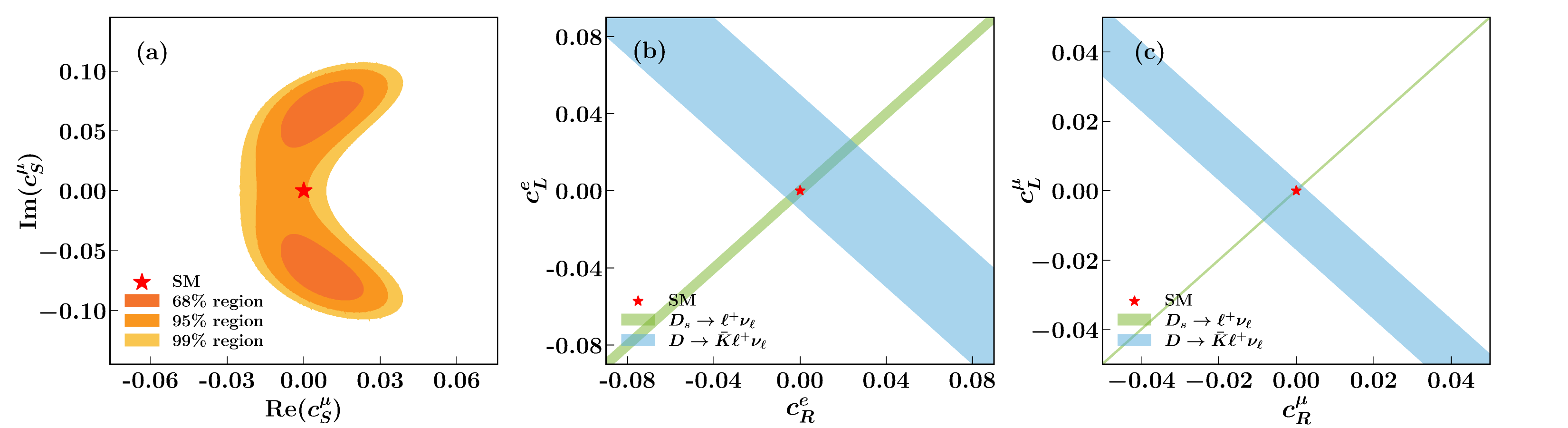}
	\caption{The confidence regions of (a)  the scalar combination of complex Wilson coefficients $c_S^{\mu}$ with different  probabilities and (b)(c) the right- and left-handed real Wilson coefficients $c_R^{\ell}$ and $c_L^{\ell}$ with  probabilities of 68\% by combining with $D_s^+\to\ell^+\nu_{\ell}$. The red dot denotes the SM predictions.}
	\label{fig:connew}
\end{figure}

By analyzing 20.3 fb$^{-1}$ of data collected at $\sqrt{s} = 3.773$ GeV with the BESIII detector, the absolute branching fractions for the decays $D^+ \to \eta e^+ \nu_e$ and $D^+ \to \eta \mu^+ \nu_\mu$ have been determined to be $(9.75 \pm 0.29 \pm 0.28) \times 10^{-4}$ and $(9.12 \pm 0.47 \pm 0.28) \times 10^{-4}$, respectively. The $\eta$ meson is reconstructed using the decay modes $\eta \to \gamma\gamma$ and $\eta \to \pi^+ \pi^- \pi^0$~\cite{BESIII:2025hjc}. The ratio $\mathcal{R}_{\mu/e}^{\eta}$ is found to be $0.94 \pm 0.06 \pm 0.03$, which is consistent with theoretical calculations~\cite{Soni:2018adu}.
From a simultaneous fit to the partial decay rates of $D^+ \to \eta \ell^+ \nu_\ell$, the product of the hadronic form factor $f_+^{D \to \eta}(0)$ and the CKM matrix element $|V_{cd}|$ is extracted to be $0.078 \pm 0.002_{\rm stat} \pm 0.001_{\rm syst}$. Using the value of $|V_{cd}|$ from the PDG~\cite{PDG:2024cfk} as input, the form factor $f_+^{D \to \eta}(0)$ is obtained as $0.374 \pm 0.007_{\rm stat} \pm 0.003_{\rm syst}$ with an improvement in precision by a factor of 3.4 compared to previous measurements~\cite{BESIII:2025hjc}.

\subsection{$D \to V \ell^+ \nu_{\ell}$, ($V= \bar{K}^{*}$)}

Using the 20.3 fb$^{-1}$ of $e^+ e^-$ dataset collected near the thresholds of $\psi(3770)$, the semeleptonic decays $D\to \bar{K}\pi\ell^+\nu_{\ell}$ are studied through a comprehensive helicity amplitude analysis. This analysis employs five independent kinematic variables: $M^2_{\bar{K} \pi}$, $q^2$, $\theta_\ell$, $\theta_K$, and $\chi$. These variables allow for the disentangling of the dominant $\bar{K}^*$ ($P$-wave) component and the $S$-wave contribution, providing a detailed understanding of the decay dynamics and the contributions from different partial waves.

The semileptonic decay $D^+ \to K_S^0 \pi^0 \mu^+ \nu_\mu$ is observed for the first time. The first complete angular analysis of $D^+ \to \bar{K}^*(892)^0 \ell^+ \nu_\ell$ is performed, yielding $CP$-averaged and $CP$-asymmetric forward-backward and triple-product asymmetries, all in agreement with SM predictions. The ratio $\mathcal{R}_{\mu/e} = 0.94 \pm 0.02{\rm stat} \pm 0.01_{\rm syst}$ is consistent with LFU, with no significant deviations across five $q^2$ bins~\cite{BESIII:2025fso}.
The $D$-wave component ($K_2^{*}(1430^{-})$) is observed for the first time in the decays $D^0 \to \bar{K}^0 \pi^- \ell^+ \nu_\ell$ and $D^0 \to K^- \pi^+ e^+ \nu_e$, with significances of 8.0$\sigma$ and 7.9$\sigma$, respectively~\cite{BESIII:2026txt,BESIII:2026ssp}. A model-independent phase shift measurement of the $S$-wave is performed with the decay $D^0 \to \bar{K} \pi \ell^+ \nu_\ell$~\cite{BESIII:2025fso}.
Fitted parameters of the amplitude analyses for the decays $D^+ \to K_S^0 \pi^0 \ell^+ \nu_\ell$~\cite{BESIII:2025fso}, $D^0 \to \bar{K}^0 \pi^- \ell^+ \nu_\ell$~\cite{BESIII:2026txt}, and $D^0 \to K^- \pi^+ e^+ \nu_e$~\cite{BESIII:2026ssp} are summarized in Table~\ref{table:kpiellnu}. Additionally, isospin channel analyses of $D^0 \to \bar{K}^*(892)^- \ell^+ \nu_\ell$ have been performed using a sub-dataset of 7.93 fb$^{-1}$ collected at 3.773 GeV~\cite{BESIII:2024xjf,BESIII:2024qnx,BESIII:2025wex}.

\begin{table}[htbp]
	\caption{The fitted parameters of the amplitude analysis, where the first uncertainty is statistical and the second is systematic, respectively.}
	\label{table:kpiellnu}
	\resizebox{\linewidth}{!}{
	\begin{tabular}{lccclc}
		\hline\hline
		Variable                                  & $D^+\to K_S^0\pi^0e^+\nu_e$ & $D^+\to K_S^0\pi^0\mu^+\nu_\mu$ & \multicolumn{2}{c}{$D^0\to \bar{K}^0\pi^-\ell^+\nu_{\ell}$} & $D^0\to K^-\pi^0 e^+\nu_e$                                 \\ \hline
		$\mathcal{B}(D\to K\pi \ell^+\nu_{\ell})$ & $(0.943\pm0.012\pm0.010)\%$ & $(0.896\pm0.017\pm0.008)\%$     & $(1.447\pm0.012\pm0.009)\%$  & $(1.391\pm0.013\pm0.008)\%$  & \multicolumn{1}{l}{$(7.878\pm0.063\pm0.048)\times10^{-3}$} \\
		$m_{\bar{K}^{*0}}$ (MeV/$c^2$)            & \multicolumn{2}{c}{$897.3\pm0.3\pm2.2$}                       & \multicolumn{2}{c}{$892.7\pm0.2$}                           & $892.9\pm0.2$                                              \\
		$\Gamma_{\bar{K}^{*0}}$ (MeV/$c^2$)       & \multicolumn{2}{c}{$45.2\pm0.6\pm0.5$}                        & \multicolumn{2}{c}{$45.6\pm0.4$}                            & $47.9\pm0.5\pm0.4$                                         \\
		$r_V$                                     & \multicolumn{2}{c}{$1.42\pm0.03\pm0.02$}                      & \multicolumn{2}{c}{$1.444\pm0.026\pm0.011$}                 & $1.41\pm0.05\pm0.01$                                       \\
		$r_2$                                     & \multicolumn{2}{c}{$0.75\pm0.03\pm0.01$}                      & \multicolumn{2}{c}{$0.752\pm0.020\pm0.004$}                 & $0.77\pm0.04\pm0.02$                                       \\
		$r_S^{(1)}$                               & --                          & --                              & \multicolumn{2}{c}{$-0.05\pm0.04\pm0.03$}                   & $0.14\pm0.04\pm0.01$                                       \\
		$a_{\rm S, BC}^{1/2}$ (GeV/$c$)$^{-1}$    & $2.32\pm0.11\pm0.25$        & $1.47\pm0.25\pm0.22$            & \multicolumn{2}{c}{$1.24\pm0.01\pm0.08$}                    & $1.98\pm0.10\pm0.13$                                       \\
		$b_{\rm S, BC}^{1/2}$ (GeV/$c$)$^{-1}$    & --                          & --                              & \multicolumn{2}{c}{$-2.71\pm0.75\pm0.71$}                   & $0.57\pm0.53\pm0.27$                                       \\
		$r_S$ (GeV)$^{-1}$                        & $-8.44\pm0.13\pm0.37$       & $-9.59\pm0.46\pm0.58$           & \multicolumn{2}{c}{$-13.21\pm0.49\pm0.36$}                  & $-7.53\pm0.22\pm0.11$                                      \\
		$f_{S-\rm wave}$ (\%)                     & $6.39\pm0.17\pm0.14$        & $7.10\pm0.68\pm0.41$            & \multicolumn{2}{c}{$5.81\pm0.19\pm0.09$}                    & $5.86\pm0.18\pm0.21$                                       \\
		$f_{P-\rm wave}$                          & $93.50\pm0.18\pm0.28$       & $92.81\pm0.67\pm0.47$           & \multicolumn{2}{c}{$94.12\pm0.19\pm0.09$}                   & $93.97\pm0.19\pm0.21$                                      \\
		$f_{D-\rm wave}$                          & --                          & --                              & \multicolumn{2}{c}{$0.092\pm0.028\pm0.018$}                 & $0.16\pm0.05\pm0.02$                                       \\ \hline\hline
	\end{tabular}
	}
\end{table}

\subsection{$D \to A \ell^+ \nu_{\ell}$, ($A=\bar{K}_1$ or $b_1(1235)$)}

\begin{figure}
	\centering
	\includegraphics[width=0.7\linewidth]{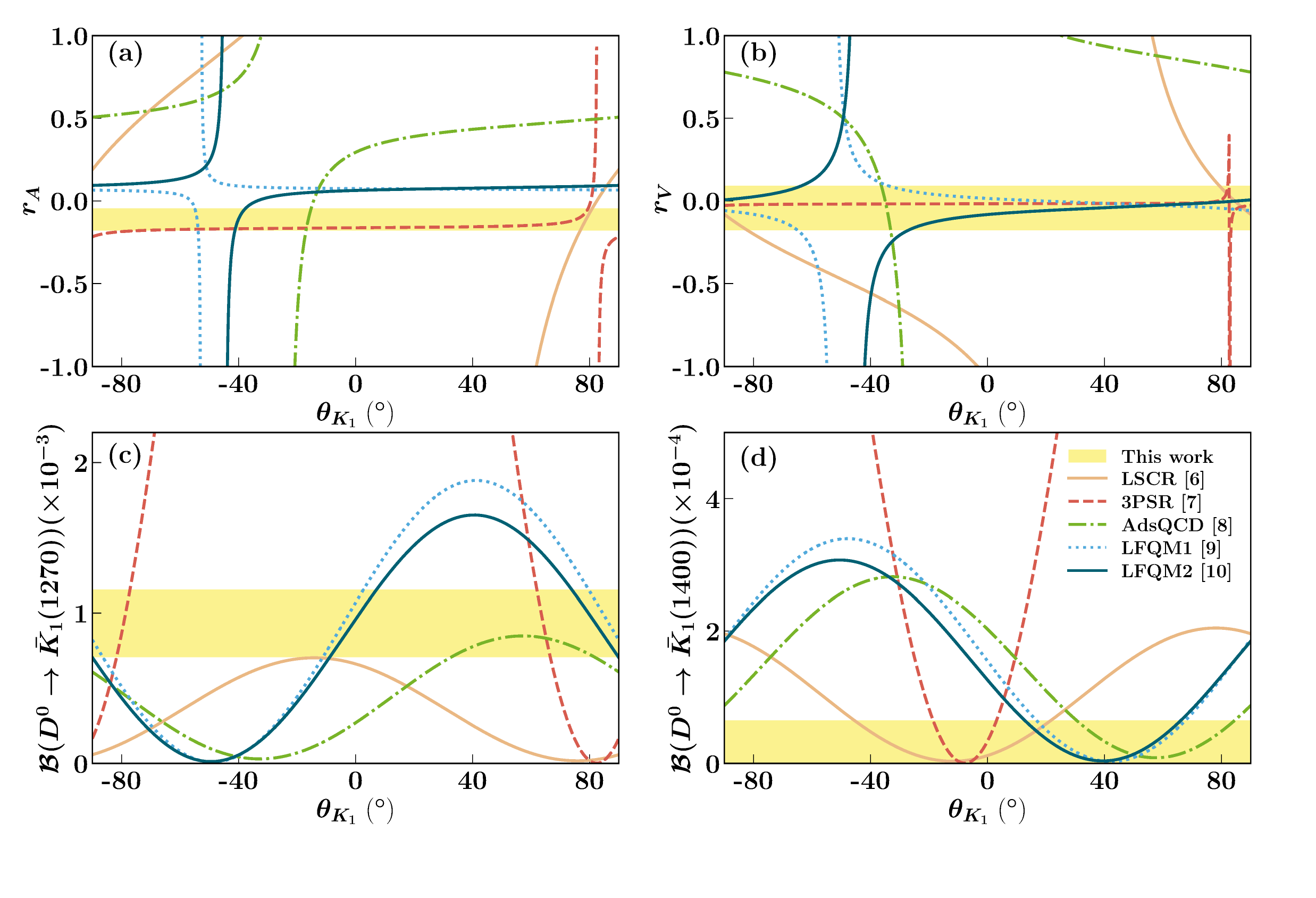}
	\caption{Comparison of (a) $r_A$, (b) $r_V$, (c) $\mathcal{B}(D^0 \to K_1(1270)^- e^+ \nu_e)$, and (d) $\mathcal{B}(D^0 \to K_1(1400)^- e^+ \nu_e)$ measured in this work and predicted by various theoretical approaches as a function of $\theta_{K_1}$.}
	\label{fig:DtoK1}
\end{figure}

Based on 20.3 fb$^{-1}$ of data at $\sqrt{s} = 3.773$ GeV, an amplitude analysis of the decays $D^{+(0)} \to K^- \pi^+ \pi^{0(-)} e^+ \nu_e$ is performed for the first time. 
The parameters of hadronic form factors for $D \to \bar{K}_1(1270)$ are determined as $r_V = V_2(0)/V_1(0) = (-4.3 \pm 1.0_{\rm stat} \pm 2.5_{\rm syst}) \times 10^{-3}$ and $r_A = A(0)/V_1(0) = (-11.2 \pm 1.0_{\rm stat} \pm 0.9_{\rm syst}) \times 10^{-3}$, while the mass and width of $\bar{K}_1(1270)$ are measured. 
The branching fractions for $D^+ \to \bar{K}_1(1270)^0 e^+ \nu_e$ and $D^0 \to \bar{K}_1(1270)^- e^+ \nu_e$ are $(2.27 \pm 0.11_{\rm stat} \pm 0.07_{\rm syst}) \times 10^{-3}$ and $(1.02 \pm 0.06_{\rm stat} \pm 0.06_{\rm syst}) \times 10^{-4}$, respectively.
No signal is found for $D^{+(0)} \to \bar{K}_1(1400) e^+ \nu_e$, and upper limits on the branching fractions are set at $1.4 \times 10^{-4}$ and $0.7 \times 10^{-4}$ for $D^+ \to \bar{K}_1(1400)^0 e^+ \nu_e$ and $D^0 \to \bar{K}_1(1400)^- e^+ \nu_e$ at 90\% confidence level.
The branching fraction ratio for $K_1(1270) \to K^* \pi$ and $K_1(1270) \to K \rho$ is $(20.3 \pm 2.1_{\rm stat} \pm 8.7_{\rm syst})$. The up-down asymmetry $\mathcal{A}'_{\rm ud} = 0.01 \pm 0.11$ and longitudinal polarization fraction $F_L = 0.50 \pm 0.04$ are consistent with SM predictions, and the mixing angle $\theta_{K_1}$ is determined from theoretical form factors which shown in Fig.~\ref{fig:DtoK1}~\cite{BESIII:2025hdt}.

Using 7.93 fb$^{-1}$ of $e^+e^-$ collision data at $\sqrt{s} = 3.773$ GeV from the BESIII detector, the decays $D^{+} \to \bar{K}_1(1270)^0 \mu^+ \nu_{\mu}$ and $D^{0} \to \bar{K}_1(1270)^- \mu^+ \nu_{\mu}$ are observed for the first time with significances of 12.5 $\sigma$ and 6.0 $\sigma$, respectively. The branching fractions are $(2.36 \pm 0.20_{-0.27}^{+0.18}) \times 10^{-3}$ for $D^+$ and $(0.78 \pm 0.11_{-0.09}^{+0.05}) \times 10^{-3}$ for $D^0$ channel~\cite{BESIII:2025yot}. Additionally, the decay $D^{0} \to b_1(1235)^{-} e^+ \nu_e$ is observed with a significance of 12.5 $\sigma$, while evidence for $D^{-} \to b_1(1235)^0 e^+ \nu_e$ is obtained with 3.1 $\sigma$ via $b_1(1235)\to\omega_{\pi^+\pi^-\pi^0}\pi$. The product branching fractions are $\mathcal{B}(D^{0} \to b_1(1235)^{-} e^+ \nu_e) \times \mathcal{B}(b_1(1235)^{-} \to \omega_\pi^{-}) = (0.72 \pm 0.18_{-0.08}^{+0.06}) \times 10^{-4}$ and $\mathcal{B}(D^{+} \to b_1(1235)^0 e^+ \nu_e) \times \mathcal{B}(b_1(1235)^0 \to \omega_\pi^0) = (1.16 \pm 0.44 \pm 0.16) \times 10^{-4}$~\cite{BESIII:2024pwp}.

\subsection{$D \to S \ell^+ \nu_{\ell}$, ($S=a_{0}(980)$)}

The branching fraction of the decay $D^0\to a_0(980)^{-}e^+\nu_e$ is measured to be $(0.86\pm0.17_{\rm stat}\pm0.05_{\rm syst})\times 10^{-4}$ with 7.93 fb$^{-1}$ of $e^+e^-$ collision data at $\sqrt{s} = 3.773$ GeV. The $f_+^{D \to a_0(980)}(0)|V_{cd}|$ is firstly determined to be $0.126\pm0.013_{\rm stat}\pm0.003_{\rm syst}$ with a single-pole parametrization of the hadronic form factor and the Flatté formula describing the $a_0(980)$ line shape in the decay dynamics analysis. Taking the $V_{cd}$ from PDG~\cite{PDG:2024cfk}, the $f_+^{D \to a_0(980)}(0)$ is extracted to be $0.559\pm0.056_{\rm stat}\pm0.013_{\rm syst}$~\cite{BESIII:2024zvp}.

\section{Summary}

\begin{figure}[htbp]
	\centering
	\includegraphics[width=0.4\linewidth]{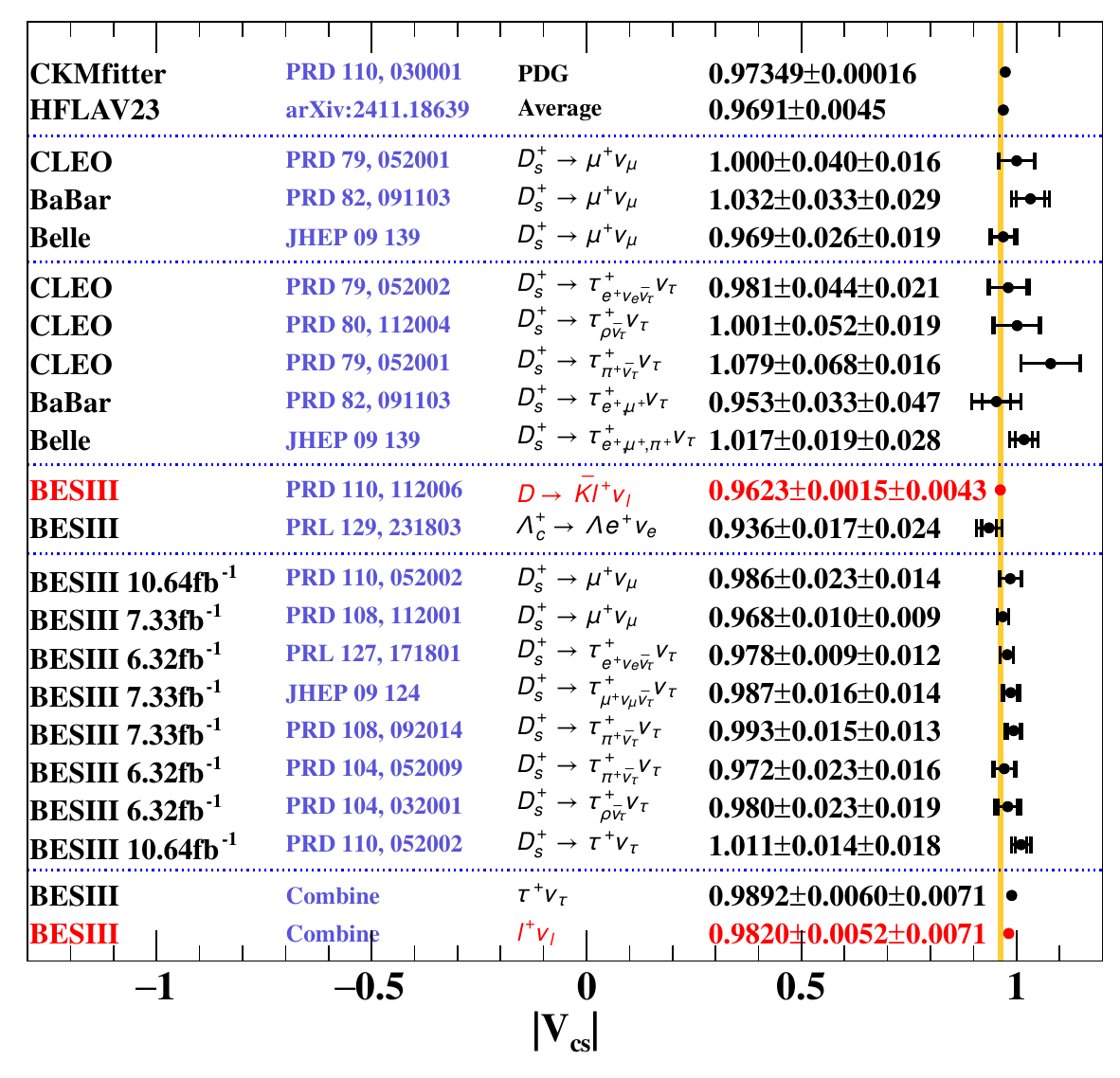}
	\includegraphics[width=0.4\linewidth]{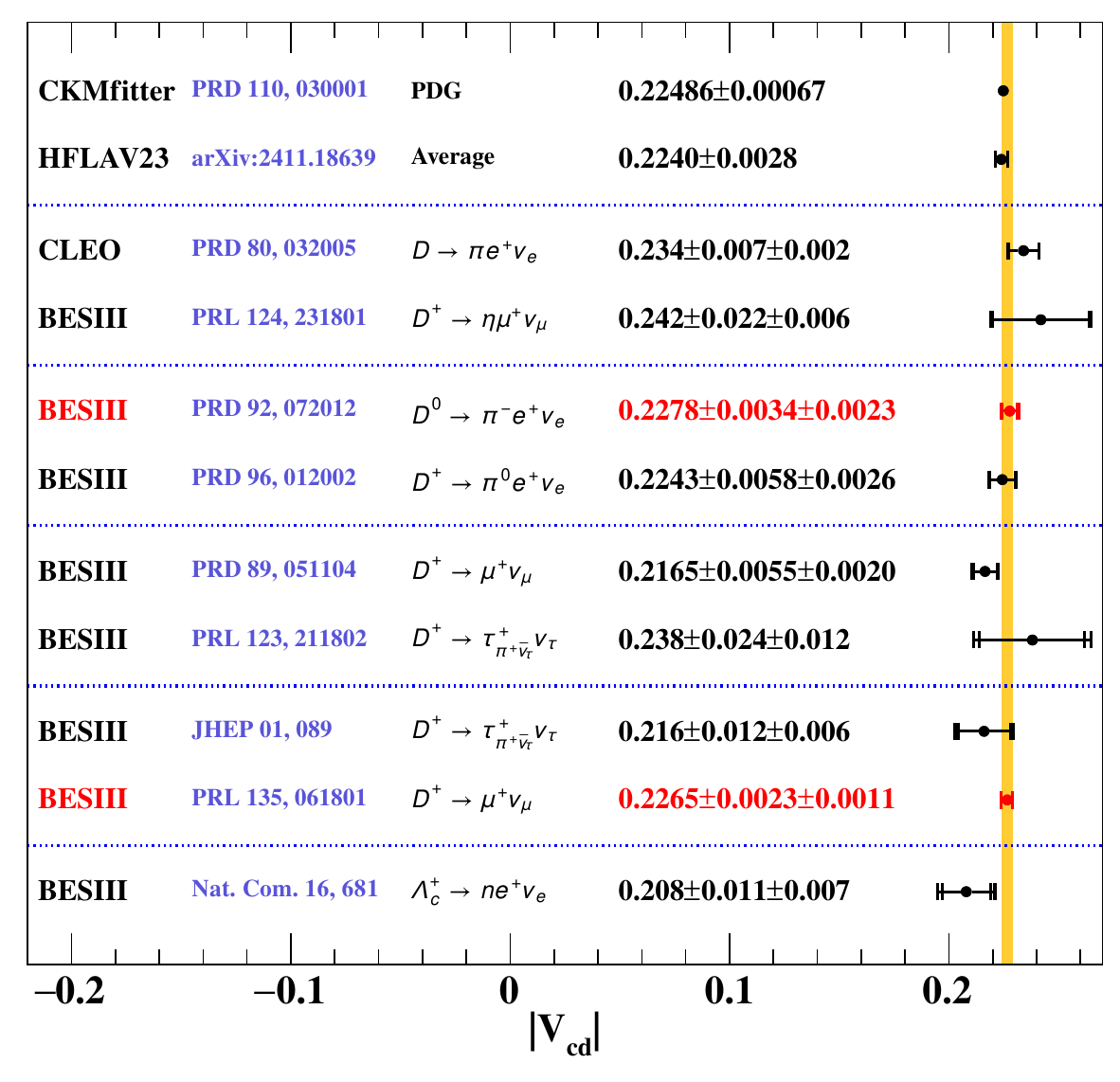}
	\caption{Comparison of the results for the CKM elements $|V_{cs}|$ and $|V_{cd}|$.}
	\label{fig:CKM}
\end{figure}

BESIII has made significant strides in precision measurements of charm leptonic and semileptonic decays across various energies. This proceeding presents the most precise single determinations of decay constants and form factors, along with rigorous tests of LFU in the charm sector. Furthermore, the CKM matrix elements $|V_{cs}|$ and $|V_{cd}|$, which are crucial for testing CKM unitarity, have been directly measured with precisions of 0.23\% and 1.2\%, respectively, as shown in Fig.~\ref{fig:CKM}. Further improvements in $D$-meson decay measurements are expected with the full 20.3 fb$^{-1}$ dataset collected near the $\psi(3770)$ threshold.

\section*{Acknowledgments}

This work is supported in part by National Key R\&D Program of China under Contracts No. 2023YFA1606000 and National Natural Science Foundation of China (NSFC) under Contracts No. 12192264.


\section*{References}

\begin{thebibliography}{99}    
	
	\bibitem{BESIII:2009fln}
	M.~Ablikim \textit{et al.} (BESIII Collaboration), {\Journal{\NIMA}{614}{345}{2010}.}
	
	
	\bibitem{Yu:2016cof}
	C.~Yu \textit{et al.} BEPCII Performance and Beam Dynamics Studies on Luminosity, in 7th
	International Particle Accelerator Conference, {TUYA01, 2016.}
	
	
	\bibitem{BESIII:2020nme}
	M.~Ablikim \textit{et al.} (BESIII Collaboration), {\Journal{\CPC}{44}{040001}{2020}.}
	
	
	\bibitem{MARK-III:1985hbd}
	R.~M.~Baltrusaitis \textit{et al.} (MARK-III Collaboration), {\Journal{\PRL}{56}{2140}{1986}.}
	
	
	\bibitem{Silverman:1988gc}
	D.~Silverman and H.~Yao, {\Journal{\PRD}{38}{214}{1988}.}
	
	
	\bibitem{BESIII:2024kvt}
	M.~Ablikim \textit{et al.} (BESIII Collaborations), {\Journal{\PRL}{135}{061801}{2025}.}
	
	
	\bibitem{PDG:2024cfk}
	S.~Navas \textit{et al.} (Particle Data Group), {\Journal{\PRD}{110}{030001}{2024}.}
	

	\bibitem{FLAG:2021npn}
	Y.~Aoki \textit{et al.} (Flavour Lattice Averaging Group (FLAG) Collaboration), {\Journal{\EPJC}{82}{869}{2022}.}

	
	\bibitem{BESIII:2021anh}
	M.~Ablikim \textit{et al.} (BESIII Collaborations), {\Journal{\PRD}{104}{052009}{2021}.}
	
	
	\bibitem{BESIII:2023cym}
	M.~Ablikim \textit{et al.} (BESIII Collaborations), {\Journal{\PRD}{108}{112001}{2023}.}
	
	
	\bibitem{BESIII:2023fhe}
	M.~Ablikim \textit{et al.} (BESIII Collaborations), {\Journal{\PRD}{108}{092014}{2023}.}
	
	
	\bibitem{BESIII:2021wwd}
	M.~Ablikim \textit{et al.} (BESIII Collaborations), {\Journal{\PRD}{104}{032001}{2021}.}
	
	
	\bibitem{BESIII:2021bdp}
	M.~Ablikim \textit{et al.} (BESIII Collaborations), {\Journal{\PRL}{127}{171801}{2021}.}
	
	
	\bibitem{BESIII:2023ukh}
	M.~Ablikim \textit{et al.} (BESIII Collaborations), {\Journal{\JHEP}{09}{124}{2023}.}
	
	
	\bibitem{BESIII:2024dvk}
	M.~Ablikim \textit{et al.} (BESIII Collaborations), {\Journal{\PRD}{110}{052002}{2024}.}
	
	
	\bibitem{FermilabLattice:2022gku}
	A.~Bazavov \textit{et al.} (Fermilab Lattice and MILC Collaborations), {\Journal{\PRD}{107}{094516}{2023}.}
	
	
	\bibitem{BESIII:2026ydr}
	M.~Ablikim \textit{et al.} (BESIII Collaborations), {{\em arXiv:}2601.21196.}
	
	\bibitem{BESIII:2026uin}
	M.~Ablikim \textit{et al.} (BESIII Collaborations), {{\em arXiv:}2601.21185.}
	
	\bibitem{BESIII:2025hjc}
	M.~Ablikim \textit{et al.} (BESIII Collaborations), {{\em arXiv:}2506.02521.}
	
	\bibitem{Soni:2018adu}
	N.~R.~Soni, M.~A.~Ivanov, J.~G.~K{\"o}rner, J.~N.~Pandya, P.~Santorelli and C.~T.~Tran, {\Journal{\PRD}{98}{114031}{2018}.}
	
	
	\bibitem{BESIII:2025fso}
	M.~Ablikim \textit{et al.} (BESIII Collaborations), {\Journal{\PRL}{135}{171801}{2025}.}
	
	
	\bibitem{BESIII:2026txt}
	M.~Ablikim \textit{et al.} (BESIII Collaborations), {{\em arXiv:}2603.04136.}
	
	
	\bibitem{BESIII:2026ssp}
	M.~Ablikim \textit{et al.} (BESIII Collaborations), {{\em arXiv:}2603.00743.}
	

	\bibitem{BESIII:2024xjf}
	M.~Ablikim \textit{et al.} (BESIII Collaborations), {\Journal{\JHEP}{03}{197}{2025}.}

	
	\bibitem{BESIII:2024qnx}
	M.~Ablikim \textit{et al.} (BESIII Collaborations), {\Journal{\PRL}{134}{011803}{2025}.}
	
	
	\bibitem{BESIII:2025wex}
	M.~Ablikim \textit{et al.} (BESIII Collaborations), {\Journal{\PRL}{135}{111803}{2025}.}
	
	
	\bibitem{BESIII:2025hdt}
	M.~Ablikim \textit{et al.} (BESIII Collaborations), {\Journal{\PRL}{135}{091801}{2025}.}
	
	
	\bibitem{BESIII:2025yot}
	M.~Ablikim \textit{et al.} (BESIII Collaborations), {\Journal{\PRD}{111}{L071101}{2025}.}
	
	
	\bibitem{BESIII:2024pwp}
	M.~Ablikim \textit{et al.} (BESIII Collaborations), {\Journal{\PRL}{136}{021801}{2026}.}
	
	\bibitem{BESIII:2024zvp}
	M.~Ablikim \textit{et al.} (BESIII Collaborations), {\Journal{\PRD}{111}{L091501}{2025}.}
	
\end{thebibliography}

\end{document}